# A New Formalism for Calculating Modal Contributions to Thermal Interface Conductance from Molecular Dynamics Simulations

**Authors:** Kiarash Gordiz[1], Asegun Henry[1,2,*]

**Affiliations:**

[1]George W. Woodruff School of Mechanical Engineering, Georgia Institute of Technology, Atlanta GA, 30332

[2]School of Materials Science and Engineering, Georgia Institute of Technology, Atlanta GA, 30332

*Correspondence to: ase@gatech.edu

**Abstract**: A new formalism for extracting the modal contributions to thermal interface conductance with full inclusion of anharmonicity is presented. The results indicate that when two materials are joined a new set of vibrational modes are required to correctly describe the transport across the interface. Among these new modes, certain classifications emerge, as most modes extend at least partially into the other material. Localized interfacial modes are also present and exhibit the highest conductance contributions on a per mode basis. The results also show that anharmonicity enables inelastic scattering at temperatures as low as 10K and inelastic processes contribute 20% of the conductance for the system studied.

**Text:**

When heat flows across an adjoining interface between two different materials there will be a temperature discontinuity at the interface. The interfacial heat flow can be written as the product of the interface conductance $G$, which is the inverse of the interfacial resistance, and the temperature difference across the interface $\Delta T$. The heat flow across the interface can be carried primarily by electrons in electrically conducting materials, but the contribution from the atomic motions is present in all materials. For solids and rigid molecules, these atomic motions correspond to vibrations around an equilibrium site, which can be decomposed into a series of eigen modes via the lattice dynamics (LD) formalism [1], and the modes have time varying amplitudes. At a given instant, by knowing the amplitudes of these eigen mode vibrations, one can sum the contributions of all the different eigen modes to recover the vibrations of each atom. The eigen modes are termed phonons in crystalline materials, since they generally correspond to propagating waves such as sound waves. However, in disordered/amorphous solids or molecules many eigen modes may not propagate or resemble the usual definition of a phonon. This has prompted our use of the term eigen modes in the ensuing discussion in order to retain generality, since the formalism presented herein is not restricted to crystalline solids.

Within the last decade, techniques have been developed to accurately calculate the individual eigen mode contributions to thermal conductivity from first principles [2-4]. These techniques now allow for predictive calculations of modal thermal conductivity for materials and nanostructures that have yet to be synthesized [5]. Knowing the contributions of specific eigen modes can enable rational design and selection of materials by crafting certain features that will target certain group of modes (i.e., acoustic, optical, longitudinal or transverse), to either inhibit or enhance their transport [6-9]. This quantitative capability has improved our ability to predict classical size effects [10] and other nanoscale phenomena [11], which are important effects that ultimately limit heat dissipation in applications such as microelectronics [12,13]. As feature dimensions in micro/nanoelectronics continue to shrink, interfacial resistance is now much more important and in some cases can become the bottleneck to heat removal [14,15]. In light of the tremendous advancements in predicting thermal conductivity over the last decade, one could argue that few deficiencies remain in our fundamental understanding of thermal conductivity. Although a few exceptions may exist [16], thermal conductivity is in principle a solved problem [2-4,17-19]. Thermal interface conductance, however, is far from being a solved problem. The central issue is that we lack quantitative understanding of the underlying processes that occur at interfaces because we currently have no way of determining the modal contributions with full inclusion of anharmonicity.

Over the last 25 years, a variety of methods have been developed [20-26], but none of the methods that provide mode level details have fully included anharmonicity. Non-equilibrium molecular dyanmics (NEMD) fully includes anharmonicity and has been used extensively to analyze different interface materials and interface qualities [27-33]. However, a formalism that can be used to study the modal contributions to $G$ in the context of NEMD is lacking. As a result, the predictive power of methods that provide mode level detail has been limited to low cryogenic temperatures, while most engineering applications involve temperatures above or near room temperature. Among the various methods that have been developed to investigate the modal contributions to $G$ [20-23,25,26,34], the most prominent are the acoustic mismatch model (AMM) [20,21], the diffuse mismatch model (DMM) [22,23,35], the atomistic Green's function (AGF) approach [26,34,36-39], and the wave packet (WP) method [25,40-44].

The AMM and DMM take the limit of purely specular and diffuse scattering respectively. Many improvements have been made to these methods [35,45-49], but neither can include the atomic level detail of the interface quality (e.g., roughness, interatomic diffusion, stress, imperfections etc.). The development of the AGF method was a major step forward, as it incorporated the atomic level details and also accounts for quantum effects [38,50]. However, most applications of the AGF method have been limited to small system sizes and harmonic interactions, due to analytical complexity and computational expense [38]. Mingo nonetheless has shown that, in principle, anharmonicity can be included in the AGF [38]. To our knowledge, however, the anharmonic AGF has not yet been widely used. The WP method, in principle, can include anharmonicity to full order, since it is a molecular dynamics (MD) based method [38,50-52]. However, in order to determine a mode's transmissivity, the WP method requires that all other modes have zero amplitude. This effectively corresponds to the $T=0K$ limit and, therefore, simply reproduces the same results as the AGF approach [14].

Each of these four methods have only been able to evaluate elastic scattering interactions where the transmission of a mode's energy across the interface is purely governed by whether or not other modes with similar frequency exist on the other side of the interface [14,39]. It has been argued that inelastic scattering may not be important at room temperature, particularly in systems with nanoscale features [26]. While others have argued that anharmonicity can have notable contributions to the thermal interface conductance at high temperatures [35,51-55] and until a method that includes anharmonicity/inelastic scattering can be tested, it is difficult to conclude whether or not anharmonicity is important. The development and testing of such a method is the central focus of the ensuing discussion.

There have been two major breakthroughs over the past decade that have enabled our new approach. First, McGaughey and Kaviany showed that by projecting the instantaneous positions of the atoms in an MD simulation onto the eigen mode solutions (i.e., the mode shapes from LD), one can calculate the instantaneous mode amplitudes as follows,

$$X(\mathbf{k},\nu) = \sum_{i=1}^{N} \frac{m_i^{1/2}}{N^{1/2}} \exp(-i\mathbf{k}\cdot\mathbf{r_i}) \mathbf{p_i}^* \cdot \mathbf{x_i} \tag{1}$$

where $\mathbf{k}$ is the wave vector, $\nu$ is the mode frequency, $\mathbf{p_i}$ is the polarization vector for atom $i$, $m_i$ is the mass for atom $i$, $\mathbf{r_i}$ is the equilibrium position for atom $i$, $\mathbf{x_i}$ is the displacement from the equilibrium for atom $i$, and $N$ is the total number of atoms. From the mode amplitude $X$, one can track the equilibrium fluctuations in the mode energy and can directly calculate the relaxation times of individual modes [3,4]. This approach agrees with other independent methods and has contributed to the ability to now calculate individual mode contributions to thermal conductivity from first principles [2,56,57]. The second major development that enabled our new formalism was introduced first by Barrat *et al.* [58] and later by Domingues *et al.* [59]. Both derived an expression for the conductance between any two groups of atoms, based on the fluctuation dissipation theorem [60]. This approach has shown agreement with other methods such as NEMD [61] and now allows for calculation of thermal interface conductance without requiring an externally applied heat flow. It is also important to note that the method introduced by Barrat *et al.* and Domingues *et al.* is general and can be applied to any phase of matter, where the atoms are simply divided into two groups such that the interface can have any arbitrary shape. Our new formalism, introduced in the next section combines the essential features of the modal decomposition method introduced by McGaughey and Kaviany with the equilibrium MD (EMD) conductance expression derived by Barrat *et al.* and Domingues *et al.*. With this new approach we can determine the modal contributions to conductance at an interface and gauge the effects of anharmonicity/inelastic scattering as well as others (i.e., roughness, interdiffusion, stress, imperfections etc.) with fidelity.

**New Formalism**

Suppose that we construct a system of two materials where each atom vibrates around an equilibrium site. The two materials can be labeled A and B and consist of $N_A$ and $N_B$ atoms respectively that can move in 3-dimensions. When we bring these two systems into contact forming an interface, we can solve the equations of motion in the limit that the interactions are harmonic to obtain a set of $3N = 3(N_A + N_B)$ eigen solutions via the LD formalism [1]. We can then write the atomic displacements and velocities as follows,

$$\mathbf{x_i} = \sum_n \frac{1}{(Nm_i)^{1/2}} \mathbf{e_{n,i}} X_n \tag{2}$$

$$\dot{\mathbf{x}}_\mathbf{i} = \sum_n \frac{1}{(Nm_i)^{1/2}} \mathbf{e_{n,i}} \dot{X}_n \tag{3}$$

where the summations are over all the eigen modes $n$ in the system, $\mathbf{x_i}$ is the displacement from equilibrium for atom $i$, $\dot{\mathbf{x}}_\mathbf{i}$ is the velocity of atom $i$, $m_i$ is the mass of atom $i$, and $\mathbf{e_{n,i}}$ is the eigen vector describing the direction and magnitude of the displacement of atom $i$, for mode $n$, $X_n$ and $\dot{X}_n$ are the normal mode coordinates for the position and velocity of mode $n$, which can be calculated from the inverse of the operations in Eqs. (2) and (3) as,

$$X_n = \sum_i \frac{m_i^{1/2}}{N^{1/2}} \mathbf{x_i} \cdot \mathbf{e}_{n,i}^* \tag{4}$$

$$\dot{X}_n = \sum_i \frac{m_i^{1/2}}{N^{1/2}} \dot{\mathbf{x}}_\mathbf{i} \cdot \mathbf{e}_{n,i}^* \tag{5}$$

where the summations are over all the atoms $i$ in the system and * denotes complex conjugate. For crystalline materials, where there is long range order and periodicity the eigen vectors $\mathbf{e}_{n,i}$ are usually expressed in terms of a polarization vector $\mathbf{p}$ and wave vector $\mathbf{k}$ as follows,

$$\mathbf{e}_{n,i} = \exp(i\mathbf{k} \cdot \mathbf{r}_{i,0}) \mathbf{e}_i \binom{\mathbf{k}}{\mathbf{p}} \tag{6}$$

where $\mathbf{r}_{i,0}$ is the equilibrium position, and $\mathbf{e}_i\binom{\mathbf{k}}{\mathbf{p}}$ is the eigen mode displacement for atom $i$. This is the more common definition, which is consistent with Eq. (1). Here, however, we have used a single index $n$ as a label for the eigen solutions instead of the wave vector $\mathbf{k}$ and polarization/frequency to identify each mode. This is done to retain generality, since it is not required that the system exhibit long range order/periodicity and therefore not all modes have to correspond to propagating wave solutions.

Following the approach pioneered by Barrat *et al.* and Dominguez *et al.*, we can write the instantaneous heat flow across the interface by simply grouping the atoms into two groups, namely A and B. At each instant, in a microcanonical ensemble, the rate at which energy is transmitted across the boundaries of material A is equal to the instantaneous rate of change of the energy in material B. The Hamiltonian of a system having N atoms can be written as,

$$H = \sum_i^N \frac{\mathbf{p}_i^2}{2m_i} + \Phi(\mathbf{r}_1, \mathbf{r}_2, \cdots, \mathbf{r}_n) \tag{7}$$

where $\mathbf{r}_i$ and $\mathbf{p}_i$ represent the position and momentum of atom $i$, respectively. The individual Hamiltonian for atom $i$ can then be written as,

$$H_i = \mathbf{p}_i^2 \big/ 2m_i + \Phi_i(\mathbf{r}_1, \mathbf{r}_2, \cdots, \mathbf{r}_n) \tag{8}$$

Using this definition for an individual atom's Hamiltonian, the instantaneous energy exchanged between material A and B is given by,

$$Q_{A \to B} = -\sum_{i \in A} \sum_{j \in B} \sum_{\alpha=x,y,z} \left\{ \frac{p_{i,\alpha}}{m_i} \left( \frac{-\partial H_j}{\partial r_{i,\alpha}} \right) + \frac{p_{j,\alpha}}{m_j} \left( \frac{\partial H_i}{\partial r_{j,\alpha}} \right) \right\} \tag{9}$$

Equation (6) is general and can be applied to any model for the atomic interactions that can be represented as a sum of individual atom energies. If only pairwise interactions are present between material A and B, Eq. (6) can be reduced to,

$$Q_{A \to B} = -\frac{1}{2} \sum_{i \in A} \sum_{j \in B} \mathbf{f}_{ij} \cdot \left( \dot{\mathbf{x}}_i + \dot{\mathbf{x}}_j \right) \tag{10}$$

where $\mathbf{f}_{ij}$ is the pairwise interaction between the two materials [54,59,62]. For pairwise interactions, it is a natural choice to partition half of the energy in the interaction with atom $i$ and the other half with atom $j$. From this relation, Domingues et al. [59] used the fluctuation dissipation theorem to show that the correlation in the equilibrium fluctuations of the heat flow are related to the conductance via,

$$G = \frac{1}{A k_B T^2} \int_0^\infty \langle Q_{A \to B}(t) \cdot Q_{A \to B}(0) \rangle dt \tag{11}$$

where $G$ is the thermal conductance between the two materials, $A$ is the interface contact area, $k_B$ is the Boltzmann constant, $T$ is the equilibrium temperature of the system, and $\langle \cdots \rangle$ represents the autocorrelation function. This result is in principle similar to the result more widely used for thermal conductivity $\kappa$ calculations,

$$\kappa = \frac{V}{k_B T^2} \int_0^\infty \langle \mathbf{Q}(t) \cdot \mathbf{Q}(0) \rangle dt \tag{12}$$

which is often referred to as the Green-Kubo formula [2,63]. In Eq. (12), $V$ is the volume of the system, and $\mathbf{Q}$ is the volume averaged heat flux [64]. To simplify the nomenclature, we will use $Q$ instead of $Q_{A \to B}$ for interfacial heat flow throughout the rest of this report.

It follows from Eq. (11) that if one could obtain the modal contributions to the heat flow across the interface such that at every instant the sum of those contributions returned the total $Q$,

$$Q = \sum_n Q_n \tag{13}$$

then $G$ can be rewritten as,

$$G = \frac{1}{A k_B T^2} \int \left\langle \sum_n Q_n(t) \cdot Q(0) \right\rangle dt = \sum_n \frac{1}{A k_B T^2} \int \langle Q_n(t) \cdot Q(0) \rangle dt \tag{14}$$

This would then give the individual contribution of each mode to G as,

$$G_n = \frac{1}{A k_B T^2} \int \langle Q_n(t) \cdot Q(0) \rangle dt \tag{15}$$

where

$$G = \sum_n G_n \tag{16}$$

The critical step is then to determine $Q_n$, subject to the requirement that $Q = \sum_n Q_n$. This can be done by replacing the velocity of each atom in Eq. (9) with the sum of modal contributions in Eq. (3),

$$Q = \sum_{i \in A} \sum_{j \in B} \sum_{\alpha=x,y,z} \left\{ \left( \sum_n \frac{1}{(Nm_i)^{1/2}} e_{n,i,\alpha} \dot{X}_n \right) \left( \frac{\partial H_j}{\partial r_{i,\alpha}} \right) - \left( \sum_n \frac{1}{(Nm_j)^{1/2}} e_{n,j,\alpha} \dot{X}_n \right) \left( \frac{\partial H_i}{\partial r_{j,\alpha}} \right) \right\}$$

$$Q_n = \frac{1}{N^{1/2}} \sum_{i \in A} \sum_{j \in B} \sum_{\alpha=x,y,z} \left\{ \left( \frac{1}{(Nm_i)^{1/2}} e_{n,i,\alpha} \dot{X}_n \right) \left( \frac{-\partial H_j}{\partial q_{i,\alpha}} \right) + \left( \frac{1}{(Nm_j)^{1/2}} e_{n,j,\alpha} \dot{X}_n \right) \left( \frac{\partial H_i}{\partial q_{j,\alpha}} \right) \right\} \quad (17)$$

Eq. (14) is general and for pairwise interactions it simplifies to,

$$Q_n = \sum_{i \in A} \sum_{j \in B} \left( \frac{-1}{2} \right) \mathbf{f}_{ij} \cdot \left( \frac{1}{(Nm_i)^{1/2}} \mathbf{e}_{n,i} \dot{X}_n + \frac{1}{(Nm_j)^{1/2}} \mathbf{e}_{n,j} \dot{X}_n \right) \quad (18)$$

The presumption in using the substitution of the modal contributions to velocity (Eq. (2)) is that any quantity that is a direct function of the atomic displacements and/or velocities can be decomposed into its modal contributions via Eqs. (2) and (3). This includes quantities such as temperature [65], pressure [66,67], entropy [68], heat capacity [69], etc. The issue with Eq. (9), however, is that it includes the vibrations from both sides of the interface. Using such an expression is inconsistent with the prevailing paradigm used to understand heat flow at solid interfaces, which is based on the phonon gas model (PGM) and the Landauer formalism [70]. In the (PGM) the net heat flow at an interface is written as [35,71],

$$q = \sum_{p_A} \left[ \frac{1}{V_A} \sum_{k_{x,A}=-k_{max}}^{k_{max}} \sum_{k_{y,A}=-k_{max}}^{k_{max}} \sum_{k_{z,A}=0}^{k_{max}} v_{z,A} \hbar \omega \tau_{AB} \left( f(\omega, T_A) - f(\omega, T_B) \right) \right] \quad (19)$$

where the summations are over different polarizations ($p$) and allowed wave vectors ($k_{x,y,z}$) in either material A or B. In Eq. (19) information from both sides is not required because the derivation of Eq. (19) employs the principle of detailed balance. Equation (19) therefore appears inconsistent with Eq. (10) because one cannot project the vibrations of atoms from side B onto the modes of side A. Such an operation is ill defined since there are no polarization vectors defined for atoms on side B if only the modes associated with side A are used to describe the heat flow. This issue can be overcome by rewriting Eq. (10) such that all of the energy associated with the interaction between an atom on side A and side B is attributed to the atom on side A. This results in [51,58],

$$Q = -\sum_{i \in A} \sum_{j \in B} \mathbf{f}_{ij} \cdot \mathbf{v_i} \quad (20)$$

which is a way of showing that Newton's third law and the principle of detailed balance are essentially the same statements. In this way, the modal decomposition of the heat flow can be written as,

$$Q_n = \sum_{i \in A} \sum_{j \in B} -\mathbf{f}_{ij} \cdot \left( \frac{1}{(Nm_i)^{1/2}} \mathbf{e_{n,i}} V_n \right) \quad (21)$$

where now the heat flow across the interface is composed of $3N_A$ contributions, as opposed to $3N$ contributions. Another option would be to simply use the modal information from the two sides assuming both are solid, as it should be noted that Eq. (9) is general and can be applied to any phases of matter. One could construct the eigen solutions for the entire system by taking the $3N_A$ solutions of the isolated system A and adding them to the $3N_B$ solutions of the isolated system B, forming 3N solutions. Here, one can overcome the issue of having to project the motions of atoms in group B onto modes described by group A by simply assigning a polarization vector $\mathbf{e} = 0$ to atoms on the B side, for solutions to the modes in A and vice versa. This set of 3N solutions could then be implemented in Eq. (9) without any modifications. A third choice would be to perform a LD calculation on the entire combined system AB. This would naturally lead to 3N solutions and can also be implemented in Eq. (9) without any modifications.

**Modal Basis Set**

We have now presented three choices for decomposing the heat flow across the interface, which lead to different physical interpretations. The choice of decomposition method is critical, as each has important implications that will be discussed further in the next section. Mathematically all of three choices satisfy the requirement of Eq. (13) ($Q = \sum_n Q_n$) and we have confirmed it is in fact the case numerically. In the following, we will refer to these three choices symbolically as {A/B}, {A+B}, and {AB}. The symbol {A/B} corresponds to using the modes obtained from a LD calculation of only one isolated side of the interface either A or B (i.e., $3N_A$ or $3N_B$ solutions) and $Q$ is determined from Eq. (21). The symbol {A+B} corresponds to using the modes obtained from a LD calculation of each isolated side of the interface and then adding them together to form $3N_A + 3N_B = 3N$ solutions. With this choice $Q$ is determined from Eq (9) Finally, the symbol {AB} corresponds to 3N solutions obtained from a LD calculation for the combined system, which contains the interface and associated interactions. The natural question that follows is whether or not they all give the same results and if not, which one is correct?

To answer this question we studied a simple interface between Lennard-Jones (LJ) solids, where the mass ($m$) and potential parameters ($\sigma$, $\varepsilon$) of side A corresponded to that of solid Argon, while side B utilized the same parameters, except that the atomic mass was increased by a factor of four ($m_B = 4m_A$). A test was then devised based on the WP method. This test naturally leads to four requirements that are based on our intuitive understanding of elastic scattering processes for a WP. For this test, a WP was constructed from a group of modes that have a narrow range of frequencies and a single polarization. The WP starts at a position far away from the interface on side A and is launched towards the interface with B [43,44]. When the WP reaches the interface it elastically scatters and a fraction of the energy is reflected and remains on side A, while the remaining energy is transmitted into a set of modes with similar frequency on side B. Using the energy modal analysis technique developed by McGaughey and Kaviany [3], as well as Eqs. (18) and (21) the mode energies and contributions to the interfacial heat flows are tracked in time for all three choices {A/B}, {A+B}, and {AB} using the equations previously introduced. Based on physical reasoning, we require the following features for the correct decomposition method:

1) As the WP approaches the interface, the total $Q$ must be zero since the atoms at the interface have not started moving yet. However, the non-zero individual components $Q_n$, should only correspond to the modes contained in the WP.

2) Since the scattering event will be purely elastic, as the WP reaches the interface, we should only observe heat flow contributions $Q_n$ associated with the original modes in the WP on side A or the modes on side B that get excited as a result of the WP scattering at the interface.

3) If we integrate $Q_n$ in time, we should see that only the modes that participate in the incoming WP or outgoing WPs contribute to the energy transfer across the interface.

4) The sum of the integrals $\sum_n \int Q_n(t)dt$ should equate to the net energy increase in side B, which is equal to the energy transmitted to the side B.

The results of this simple test showed that all three decomposition choices satisfy requirements 1) and 4). However, Fig. 1 shows that {A/B} and {A+B} do not satisfy requirements 2) or 3), as they both show frequency broadening when the WP reaches the interface. This seems unphysical for two reasons. First, the frequency content of all the atomic motions before, during, and after the scattering event all lie within the same frequency range as the original WP. Therefore the broadening exhibited by {A/B} and {A+B} is not representative of actual excitation of those modes, since those frequencies do not actually manifest in the simulation. Secondly, for {A/B} and {A+B}, the modes that were not initially excited in WP comprise 50% of the total transmitted energy to the other side. This also seems unphysical, since these modes were not excited at any point during the WP scattering event, yet choices {A/B} and {A+B} would show that they are responsible for half of the heat transfer. When the combined system {AB} is used, all four of the requirements are satisfied and thus, we believe {AB} is the correct basis set for decomposing interfacial heat flow.

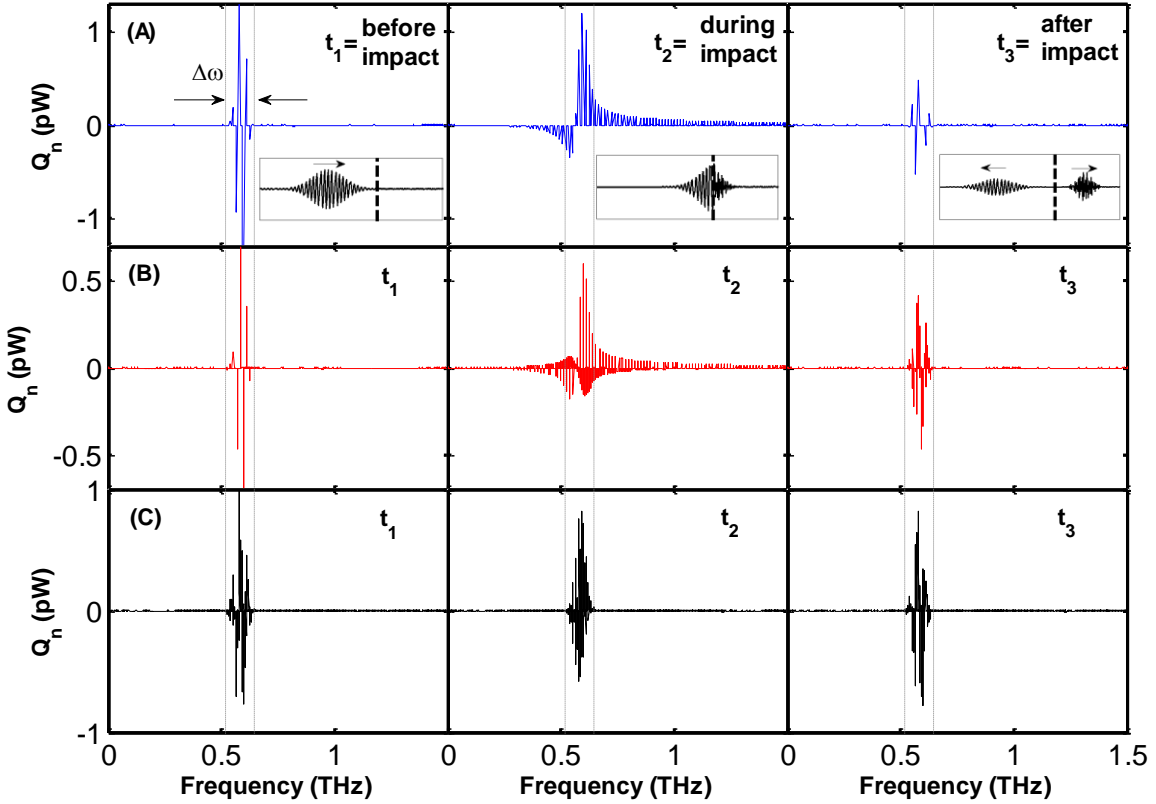

**Fig. 1.** Calculation of the modal contributions to interfacial heat flow using the three different basis sets. Figures (A), (B) and (C) show the results for {A/B}, {A+B}, and {AB}, respectively. The data is shown at three different instants during the simulation: before the impact ($t_1$), during the impact ($t_2$), and after the impact ($t_3$). The atomic displacement profiles for the WP simulation at these three times are shown as insets in (A). When the WP encounters the interface at $t_2$, the {A/B} and {A+B} basis sets show broadening, suggesting that the contributions from frequencies that are not actually present in the simulation are significant. However, the {AB} basis set calculates only the frequencies that were initially excited in the system ($\Delta\omega$ range). Therefore, only {AB} satisfies requirements 2) and 3).

LD of the combined system {AB} also reveals a key feature that is not captured by {A/B} or {A+B}, which is that atoms near the interface on side B can vibrate at frequencies above the maximum frequency allowed in the isolated material B ($\omega_{B,\max}$). As isolated systems, atoms on side A can exhibit higher frequency vibrations than atoms on side B. However, when the two systems are coupled together the heavier atoms on the B side can experience vibrations at frequencies above what is achievable in the isolated material B (e.g., $\omega_{B,\max}$). This feature is critical as it has been observed in actual MD simulations of interfaces [49,72], but cannot be captured by the conventional paradigm of {A/B}, nor {A+B}. The fact that {AB} is the only decomposition that yields correct results has important implications for the physical picture that is often used to interpret interfacial heat flow. Until now, the physical picture used to understand interfacial heat flow (e.g., interface conductance) has been the same as the physical picture used to describe heat flow through crystalline solids (e.g., thermal conductivity), namely the PGM. The PGM, however, relies on the idea that all of the eigen modes in a material correspond to propagating solutions, which lead to well defined phonon velocities. This then leads to an expression for the interfacial heat flow similar to Eq. (19). Using the {AB} modal basis set, however, is inconsistent with this physical picture, since it in no way strictly requires all modes to propagate. As a result, usage of the {AB} decomposition requires changing our physical picture of interfacial heat flow and it is one of the most important results reported herein.

Even more insight can be gained by further examination of the modes of the combined {AB} system. The {AB} system results in modes that differ from the {A/B} or {A+B} descriptions. For the systems studied herein, namely LJ solids, we have tentatively classified the 3N solutions into 4 seemingly distinct categories: <1> extended modes, <2> partially extended modes, <3> isolated modes and <4> interfacial modes. Further investigation into a wide variety of material classes is needed to determine the extent to which such classifications of modes are general. To our knowledge, this is the first report of such LD calculations, so we have only adopted this nomenclature to better communicate the results discussed herein.

Figure 2 shows one example of each of the four types of modes as well as their respective contributions to the density of states. The LD calculation results in some modes where the atomic vibrations are present throughout the entire system, and are therefore termed extended modes (e.g., type <1>). For the LJ system under study, extended modes extend from the bulk of A through the interface to the bulk of B (inset of Fig. 2A). In this sense, these modes do not actually encounter the interface and behave like long wavelength phonons/sound waves, which are largely unobstructed by the presence of the interface [26,44]. Furthermore, because both sides (A & B) vibrate at one frequency for these modes, the density of states for extended modes has a sharp cutoff at $\omega_{B,\max}$, which is the maximum frequency for the bulk portion of the heavier side (Fig. 2A). Other eigen solutions exist where all the atoms on one side of the interface vibrate and partially extend to the other side, but the vibrations do not extend through the entirety of the other side. We refer to such eigen modes as partially extended modes denoted by <2> (inset of Fig. 2B). These modes comprise the majority of the eigen solutions (see Fig. 2B). In contrast, there are other modes where the atomic vibrations are restricted to only one of the materials on one side of the interface, but there are no vibrations near the interface. These modes are termed isolated modes, denoted by <3>. In these modes, the vibrations on one side of the interface decay quickly before reaching the interface (inset of Fig. 2C). Lastly, the LD calculation also results in eigen solutions, that are localized around the interface. These modes are termed interfacial modes and are denoted by <4>. Although the density of states of the interfacial modes is much smaller than the other types of modes (Fig. 2D), we show in the next section that they play a significant role in the transfer of heat across the interface. We've also tested such modes by starting a simulation with them as a singly excited mode. Interestingly, the interfacial modes do not immediately decay and couple to other modes, as they are in fact eigen solutions to the equations of motion and are not evanescent decaying modes that immediately dissipate their energy into other modes.

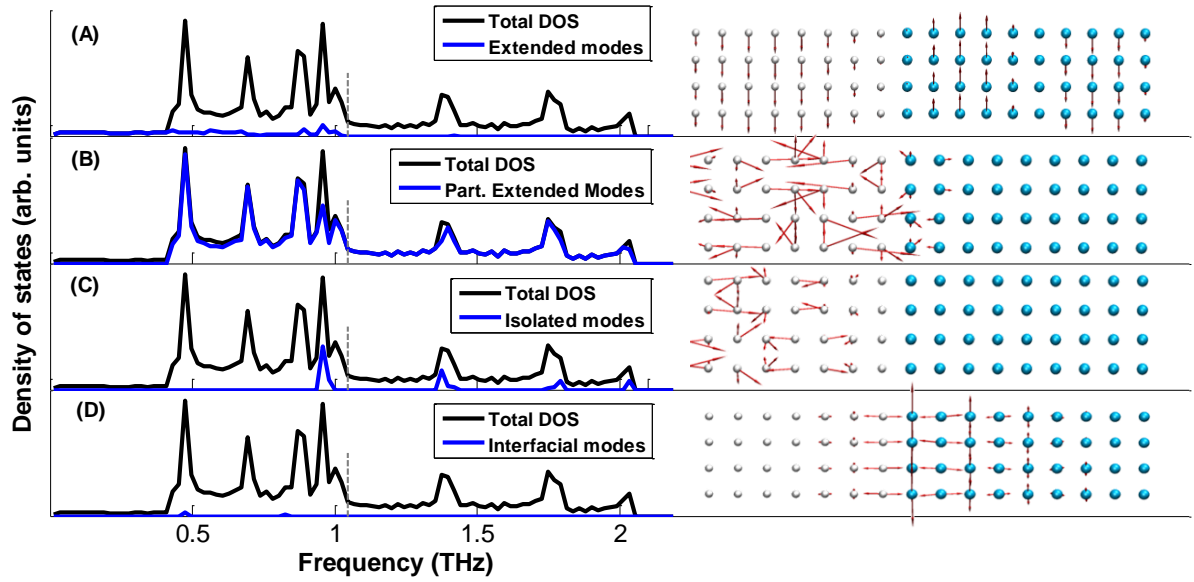

**Fig. 2.** Density of states for the four classifications of eigen modes identified for the {AB} basis set: (A) extended <1>, (B) partially extended <2>, (C) isolated <3>, and (D) interfacial <4> modes. Each inset shows the eigen vector displacements for an example of each type of solution.

**Results and Discussion**

With the correct choice of modes now clear {AB}, we conducted EMD simulations of the LJ system and obtained the modal contributions to the conductance. The conductance accumulations with respect to the eigen mode frequencies are shown in Fig. 3 for an EMD simulation at 60K. In Fig. 3, for the first time, one can clearly see the effects of anharmonicity. Elastic scattering requires that only modes with frequencies below $\omega_{B,max}$ can contribute to the conductance, because it requires that modes with similar frequency exist on the opposite side to exchange energy with. Therefore, all previous approaches reach 100% of the conductance by $\omega_{B,max}$, and all accumulation contributions above in Fig. 3 from the model described herein are a result of inelastic scattering. In our new formalism, mode conversion via inelastic scattering is possible, which allows modes above $\omega_{B,max}$ to contribute substantially to the conductance (i.e., ~ 20-25%).

Another feature of the formalism described herein is that, since Eq. (17) describes the instantaneous modal contributions to the heat flux across the interface, it can also be implemented in a NEMD simulation. Using NEMD and the {AB} basis set, we also calculated the modal contributions to conductance and show the accumulation with respect to mode frequency in Fig. 3. The EMD and NEMD results in Fig. 3 exhibit the same features, which is further confirmation of the method's validity. Although the WP test established that only the {AB} choice of basis set is correct, for comparison, Fig. 3 shows the conductance accumulations for the {A/B} and {A+B} basis sets as well. Here, it is clear that each choice exhibits distinctly different features and each results in a different contribution attributed to inelastic scattering. This difference is important, as it indicates that the three choices are not equivalent and the validity of the {AB} basis set as established by the WP test further suggests that the PGM is inconsistent with the actual dynamics at an interface.

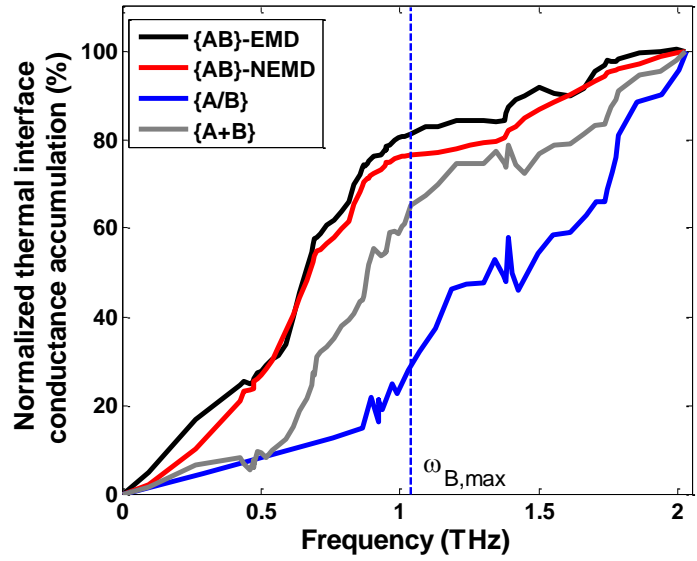

**Fig. 3.** Normalized thermal interface conductance accumulation for different basis sets at T=60K. Inelastic scattering arises from the modes with frequencies above the maximum frequency of the heavier side $\omega_{B,max}$, as indicated by the dashed line.

Using the {AB} modal basis set, we calculated the contributions of different types of modes to thermal interface conductance. The results are shown in Table 1 and indicate that interfacial modes, despite their low population, have the highest contribution on a per mode basis. This type of insight immediately leads one to wonder if the interface conductance can be increased by creating more of such modes at the interface. Presumably, the number of interfacial modes can be increased through interdiffusion at the interface, which could potentially increase the conductance [27,51]. This approach, however, may also serve to decrease the number of extended or partially extended modes or impede their propagation. Many studies to understand such tradeoffs are warranted and are enabled by this new formalism.

**Table 1**. Number of states and contributions of the eigen modes described by {AB}.

| Mode Type | Fraction of total number of states (%) | Contribution to G (%) | Contribution to G/ fraction of total number of states |
|---|---|---|---|
| <1> | 10.73 | 42.89 | 3.99 |
| <2> | 83.19 | 53.4 | 0.64 |
| <3> | 5.76 | 0.55 | 0.095 |
| <4> | 0.31 | 3.16 | 10.19 |

Another important test for the formalism described herein is the behavior at low temperatures. Although Fig. 3 shows significant effects from inelastic scattering, at sufficiently low temperatures, such interactions should cease to exist, resulting in the conductance reaching 100% by $\omega_{B,max}$. This serves as yet another test of the validity of the formalism presented herein. Using different basis sets, we have calculated modal contributions to conductance at a temperature equal to 1K (Fig. 4). At this lower temperature, the {AB} basis set results in zero contribution to the total conductance for frequencies larger than $\omega_{B,max}$. This serves as another confirmation of the formalism's validity and the correctness of the {AB} basis set. The {A/B} and {A+B} basis sets, on the other hand, still show considerable contributions for frequencies above $\omega_{B,max}$, which is not in agreement with previous models such as the AGF.

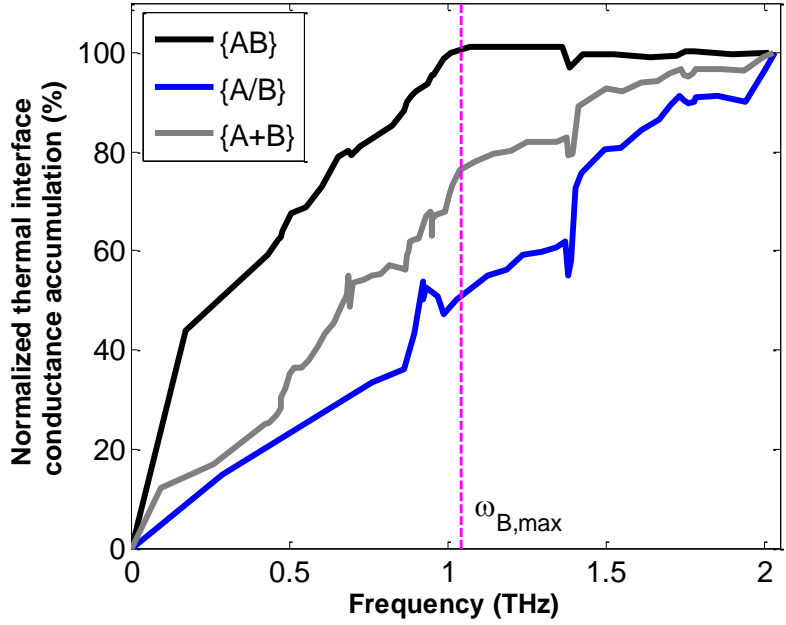

**Fig. 4.** Normalized thermal interface conductance accumulation for different basis sets at T=1K. Using the {AB} basis set, conductance contributions from frequencies above $\omega_{B,max}$ are zero.

# Conclusion

A new formalism for calculating the modal contributions to thermal interface conductance has been presented. The approach is based on the modal decomposition of the instantaneous heat flow across an interface, which can be implemented in either EMD or NEMD simulations. Results of both EMD and NEMD simulations of a LJ solid interface show agreement using the formalism presented, when the thermal interface conductance accumulation is calculated. A WP based test was used to compare the behavior of different choices for the modal decomposition. The results suggested that LD results for the combined system are needed to accurately describe the behavior at an interface, instead of using the modes of each isolated material, which is the current and prevailing paradigm based on the PGM. Low temperature calculations further validated the accuracy of the formalism and specifically the importance of using the modes of the combined system, rather than the modes of each isolated material. The usage of this basis set, which includes the presence of interfacial modes has far reaching implications for mode-mode interactions, as well as mode-electron interactions. This new perspective can provide a full picture of the modal contributions to conductance at interfaces, including anharmonicity to full order. Many additional studies to understand the effects of interdiffusion, roughness, imperfections, stress, changes in crystal structure etc. are needed. The new formalism presented herein serves as the critical step forward, which enables a high degree of fidelity in studying such effects. This new formalism can enable new understanding of the physics of energy transport at interfaces and will ultimately guide the design and selection of materials for various applications.

**Acknowledgments:** We acknowledge useful discussions with professor Michael Leamy.